%% file: main_1.tex
\documentclass[journal,12pt,onecolumn]{IEEEtran}

\usepackage{setspace}
\ifCLASSOPTIONonecolumn \doublespace \fi
\usepackage{graphics}
\usepackage{rotating}
\usepackage{amsfonts}
\usepackage{amsmath}
\usepackage{amsthm}
\usepackage{subfigure}
\usepackage{multirow}
\allowdisplaybreaks[1]
\newtheorem*{theorem}{Theorem}
\renewcommand{\IEEEQED}{\IEEEQEDopen}

\graphicspath{{Figures/}} \DeclareGraphicsExtensions{.eps,.ps}

\title{Reduced Complexity Sphere Decoding}

\ifCLASSOPTIONonecolumn
\author{
Boyu Li and Ender Ayanoglu\\
Center for Pervasive Communications and Computing\\ 
Department of Electrical Engineering and Computer Science\\
The Henry Samueli School of Engineering\\
University of California - Irvine\\
Irvine, California 92697-2625\\
Email: boyul@uci.edu, ayanoglu@uci.edu}
\date{}

\begin{document}
\maketitle







\input{Abstract}





\input{Introduction}
\input{CSD}
\input{PSD_1}
\input{Soft}
\input{Results_1}
\input{Conclusions}


\bibliographystyle{IEEEtran}
\bibliography{IEEEabrv,Mybib}
\end{document}

%% file: Abstract.tex
\begin{abstract}
In Multiple-Input Multiple-Output (MIMO) systems, Sphere Decoding (SD) can achieve performance equivalent to full search Maximum Likelihood (ML) decoding, with reduced complexity. Several researchers reported techniques that reduce the complexity of SD further. In this paper, a new technique is introduced which decreases the computational complexity of SD substantially, without sacrificing performance. The reduction is accomplished by deconstructing the decoding metric to decrease the number of computations and exploiting the structure of a lattice representation. Furthermore, an application of SD, employing a proposed smart implementation with very low computational complexity is introduced. This application calculates the soft bit metrics of a bit-interleaved convolutional-coded MIMO system in an efficient manner. Based on the reduced complexity SD, the proposed smart implementation employs the initial radius acquired by Zero-Forcing Decision Feedback Equalization (ZF-DFE) which ensures no empty spheres. Other than that, a technique of a particular data structure is also incorporated to efficiently reduce the number of executions carried out by SD. Simulation results show that these approaches achieve substantial gains in terms of the computational complexity for both uncoded and coded MIMO systems. 



\end{abstract}

%% file: Introduction.tex
\section{Introduction} \label{sec:Introduction}
Multiple-Input Multiple-Output (MIMO) systems have drawn substantial research and development interest because they offer high spectral efficiency and performance in a given bandwidth. In such systems, the goal is to minimize the Bit Error Rate (BER) for a given Signal-to-Noise Ratio (SNR). A number of different MIMO systems exist. The optimum decoding of these systems is typically highly complicated. Therefore, a number of decoding algorithms with different complexity-performance tradeoffs have been introduced. Linear detection methods such as Zero-Forcing (ZF) or Minimum Mean Squared Error (MMSE) provide linear complexity, however their performance are suboptimal. Ordered successive interference cancellation decoders such as Vertical Bell Laboratories Layered Space-Time (V-BLAST) algorithm, show slightly better performance compared to ZF and MMSE, but suffer from error propagation and are still suboptimal \cite{Burg_VLSI_SDA}. It is well-known that, for a MIMO system, Maximum Likelihood (ML) detection is optimum. However, the complexity of the ML algorithm in MIMO systems increases exponentially with the number of possible constellation points for the modulation scheme, making the algorithm unsuitable for practical purposes \cite{Zimm_SD}. Sphere Decoding (SD), on the other hand, is proposed as an alternative for ML that provides optimal or near-optimal performance with reduced complexity \cite{Jalden_SD}.

Although the complexity of SD is much smaller than ML decoding, there is still room for complexity reduction in conventional SD. To that end, several complexity reduction techniques for SD have been proposed. In \cite{Cheng_IR_SD} and \cite{Han_SD_IR}, attention is drawn to initial radius selection strategy, since an inappropriate initial radius can result in either a large number of lattice points to be searched or a large number of restart actions. In \cite{Hassibi_SD} and \cite{Zhao_SDA_IRS}, this complexity is attacked by making a proper choice to update the sphere radius. In \cite{Damen_MLD}, the Schnorr-Euchner (SE) strategy is applied to SD, which executes intelligent enumeration of candidate symbols at each level to reduce the number of visited nodes when the system dimension is small \cite{Rekaya_MLLD}. Channel reordering techniques can also be applied to reduce the number of visited nodes \cite{Rekaya_MLLD}, \cite{Azzam_SD_NLR}, \cite{Azzam_SD_RLR}. Other methods, such as the K-best lattice decoder \cite{Wong_K_Best}, \cite{Huynh_TLS_SDA}, can significantly reduce the complexity at low SNR, but with the tradeoff of BER performance degradation.

In this paper, the complexity of SD is efficiently improved by reducing the number of operations required at each node to obtain the ML solution for flat fading channels. This complexity reduction is achieved by deconstructing the decoding metric in order to reduce the number of computations and exploiting the structure of a lattice representation of SD \cite{Azzam_SD_NLR}, \cite{Azzam_SD_RLR}. In simulations, $2\times2$ and $4\times4$ MIMO systems with $4$-QAM and $64$-QAM have been studied. In these systems, the reduction in the number of real additions is in the range of $40\%-75\%$, and the reduction in the number of real multiplications is in the range of $70\%-90\%$, without any change in performance. The complexity gains increase with the MIMO system dimension or the modulation alphabet size. Moreover, for calculating the soft bit metrics of a bit-interleaved convolutional-coded MIMO system, an application of SD, employing a proposed smart implementation with very low computational complexity is also introduced. Other than the operation reduction at each node achieved by the reduced complexity SD, the initial radius of SD is acquired by Zero-Forcing Decision Feedback Equalization (ZF-DFE) \cite{Han_SD_IR}, which ensures no empty spheres. Furthermore, a technique of a particular data structure is applied to efficiently reduce the number of executions carried out by SD. Simulation results show that conventional SD substantially reduces the complexity, in terms of the average number of real multiplications needed to acquire one soft bit metric, compared with exhaustive search. With the proposed smart implementation, further reductions of orders of magnitude are achieved. As with the previous techniques, the reduction becomes larger as the MIMO system dimension or the constellation size increases. 

The remainder of this paper is organized as follows: In Section \ref{sec:CSD}, the problem definition is introduced and a brief review of conventional SD algorithm is presented. In Section \ref{sec:PSD}, a new technique to implement the SD algorithm with low computational complexity is proposed, and the mathematical derivations for the complexity reduction are carried out. In section \ref{sec:Soft}, an application of SD, employing a proposed smart implementation with very low complexity for calculating the soft bit metrics of a bit-interleaved convolutional-coded MIMO system is presented. In Section \ref{sec:Results}, complexity comparisons with different number of antennas and modulation schemes of both uncoded and coded MIMO systems are provided. Finally, a conclusion is provided in Section \ref{sec:Conclusions}.

%% file: CSD.tex
\section{Conventional Sphere Decoder} \label{sec:CSD}
In this paper, MIMO systems using square Quadrature Amplitude Modulation (QAM) with $N_t$ transmit and $N_r$ receive antennas are considered, and the channel is assumed to be flat fading. Then, the input-output relation is given by
\begin{equation}
{\tilde{\mathbf y}}=\tilde{\mathbf H}\tilde{\mathbf x}+\tilde{\mathbf n}, \label{eq:io_complex}
\end{equation}
where $\tilde{\mathbf y}\in\mathbb{C}^{N_r}$ is the $N_r$ dimensional received vector symbol and $\mathbb{C}$ denotes the set of complex numbers, $\tilde{\mathbf H}\in\mathbb{C}^{N_r{\times}N_t}$ is the channel matrix whose channel coefficients are independent and identically distributed (i.i.d.) zero-mean unit-variance complex Gaussian random variables, $\tilde{\mathbf x}\in\mathbb{C}^{N_t}$ is an $N_t$ dimensional transmitted complex vector with each element in square QAM format, and $\tilde{\mathbf n}\in\mathbb{C}^{N_r}$ is a zero-mean complex white Gaussian noise vector with variance $\sigma^2$ for each element.

Assuming $\tilde{\mathbf H}$ is known at the receiver, ML detection is
\begin{equation}
\hat{\tilde{\mathbf x}}=\arg \min_{\tilde{\mathbf x}\in\tilde{\chi}^{N_t}}\Vert \tilde{\mathbf y}-\tilde{\mathbf H}\tilde{\mathbf x}\Vert^2 \label{eq:mld}
\end{equation}
where $\tilde{\chi}$ denotes the sample space for QAM modulation scalar symbols. For example,  $\tilde{\chi}=\left\lbrace -3, -1, 1, 3\right\rbrace ^2$ for $16$-QAM.

Solving (\ref{eq:mld}) is known to be NP-hard, given that a full search over the entire lattice space is performed \cite{Hassibi_EC_ILSP}. SD, on the other hand, solves (\ref{eq:mld}) by searching only lattice points that lie inside a sphere of radius $\delta$ centering around the received vector $\tilde{\mathbf y}$. 

A frequently used solution for the QAM-modulated signal model is to decompose the $N_r$-dimensional complex-valued problem (\ref{eq:io_complex}) into a $2N_r$-dimensional real-valued problem, which can be written as
\begin{equation}
\begin{bmatrix}
\Re \left\lbrace \tilde{\mathbf y} \right\rbrace \\
\Im \left\lbrace \tilde{\mathbf y} \right\rbrace  
\end{bmatrix}
=
\setlength{\arraycolsep}{0.0em}
\begin{bmatrix}
\Re \left\lbrace \tilde{\mathbf H} \right\rbrace & -\Im \left\lbrace \tilde{\mathbf H} \right\rbrace  \\
\Im \left\lbrace \tilde{\mathbf H} \right\rbrace & \Re \left\lbrace \tilde{\mathbf H} \right\rbrace
\end{bmatrix}
\begin{bmatrix}
\Re \left\lbrace \tilde{\mathbf x} \right\rbrace \\
\Im \left\lbrace \tilde{\mathbf x} \right\rbrace  
\end{bmatrix}
+
\begin{bmatrix}
\Re \left\lbrace \tilde{\mathbf n} \right\rbrace \\
\Im \left\lbrace \tilde{\mathbf n} \right\rbrace  
\end{bmatrix},
\label{eq:io_c2r}
\end{equation}
where $\Re \left\lbrace {\mathbf r} \right\rbrace$ and $\Im \left\lbrace {\mathbf r} \right\rbrace$ denote the real and imaginary parts of ${\mathbf r}$ respectively \cite{Jalden_SD}, \cite{Hassibi_EC_ILSP}. Let
\begin{equation}
{\mathbf y}=
\begin{bmatrix}
\Re \left\lbrace \tilde{\mathbf y} \right\rbrace^T &  \Im \left\lbrace \tilde{\mathbf y} \right\rbrace^T  
\end{bmatrix}^{T},
\label{eq:receive_c2r}
\end{equation}
\begin{equation}
{\mathbf H}=
\begin{bmatrix}
\Re \left\lbrace \tilde{\mathbf H} \right\rbrace & -\Im \left\lbrace \tilde{\mathbf H} \right\rbrace  \\
\Im \left\lbrace \tilde{\mathbf H} \right\rbrace & \Re \left\lbrace \tilde{\mathbf H} \right\rbrace
\end{bmatrix},
\label{eq:channel_c2r}
\end{equation}
\begin{equation}
{\mathbf x}=
\begin{bmatrix}
\Re \left\lbrace \tilde{\mathbf x} \right\rbrace^T & \Im \left\lbrace \tilde{\mathbf x} \right\rbrace^T  
\end{bmatrix}^{T},
\label{eq:transmit_c2r}
\end{equation}
\begin{equation}
{\mathbf n}=
\begin{bmatrix}
\Re \left\lbrace \tilde{\mathbf n} \right\rbrace^T & \Im \left\lbrace \tilde{\mathbf n} \right\rbrace^T  
\end{bmatrix}^{T},
\label{eq:noise_c2r}
\end{equation}
then (\ref{eq:io_c2r}) can be written as
\begin{equation}
{\mathbf y}={\mathbf H}{\mathbf x}+{\mathbf n}. \label{eq:io_real}
\end{equation}

Assuming $N_t=N_r=N$ in the sequel, and using the QR decomposition of ${\mathbf{H}}=\mathbf{QR}$, where ${\mathbf R}$ is an upper triangular matrix, and the matrix ${\mathbf Q}$ is unitary, SD solves 
\begin{equation}
\hat{\mathbf x}=\arg \min_{{\mathbf x}\in\Omega}\Vert \breve{\mathbf y}-\mathbf{Rx}\Vert^{2}
\label{eq:sd}
\end{equation}
with $\breve{\mathbf y}={\mathbf Q}^{H}{\mathbf y}$, where $\Omega$ denotes a subset of $\chi^{2N}$ whose elements satisfy $\Vert \breve{\mathbf y}-\mathbf{Rx}\Vert^{2}<\delta^{2}$, and $\chi$ denotes the sample space for one dimension of QAM-modulated symbols, e.g., $\chi=\left\lbrace -3,-1,1,3\right\rbrace$ for $16$-QAM. 

The SD algorithm can be viewed as a pruning algorithm on a tree of depth $2N$, whose branches correspond to elements drawn from the set $\chi$ \cite{Azzam_SD_NLR}, \cite{Azzam_SD_RLR}, \cite{Hassibi_EC_ILSP}. Conventional SD implements a Depth-First Search (DFS) strategy in the tree, which can achieve ML performance. 

Conventional SD starts the search process from the root of the tree, and then searches down along branches until the total weight of a node exceeds the square of the sphere radius, $\delta^{2}$. At this point, the corresponding branch is pruned, and any path passing through that node is declared as improbable for a candidate solution. Then the algorithm backtracks and proceeds down a different branch. Whenever a valid lattice point at the bottom level of the tree is found within the sphere, $\delta^{2}$ is set to the newly-found point weight, thus reducing the search space for finding other candidate solutions. In the end, the path from the root to the leaf that is inside the sphere with the lowest weight is chosen to be the estimated solution $\hat{\mathbf x}$. If no candidate solutions can be found, the tree is searched again with a larger initial radius. 

%% file: PSD_1.tex
\section{Proposed Sphere Decoding} \label{sec:PSD}

The complexity of SD is measured by the number of operations required per visited node multiplied by the number of visited nodes throughout the search procedure \cite{Hassibi_EC_ILSP}. The complexity can be reduced by either reducing the number of visited nodes or the number of operations to be carried out at each node, or both. Making a judicious choice of initial radius to start the algorithm with \cite{Cheng_IR_SD}, \cite{Han_SD_IR}, executing a proper sphere radius update strategy \cite{Hassibi_SD}, applying an improved search strategy \cite{Damen_MLD}, and exploiting channel reordering \cite{Rekaya_MLLD}, \cite{Azzam_SD_NLR}, \cite{Azzam_SD_RLR} can all reduce the number of visited nodes. In this paper, our focus is on reducing the average number of operations required at each node for SD.

The node weight is given by \cite{Azzam_SD_NLR}, \cite{Azzam_SD_RLR},
\begin{equation}
w({\mathbf x}^{(u)})=w({\mathbf x}^{(u+1)})+w_{pw}({\mathbf x}^{(u)}), \label{eq:node_weight}
\end{equation}
for $u=2N, \cdots, 1$, with $w({\mathbf x}^{(2N+1)})=0$ and $w_{pw}({\mathbf x}^{(2N+1)})=0$, where ${\mathbf x}^{(u)}$ denotes the partial vector symbol at layer $u$. The partial weight corresponding to ${\mathbf x}^{(u)}$ is written as
\begin{equation}
w_{pw}({\mathbf x}^{(u)})=|\breve{y}_u-\sum^{2N}_{v=u}{r_{u,v}x_v}|^{2}, \label{eq:node_pw}
\end{equation}
where $r_{u,v}$ denotes the $(u,v)^{th}$ element of $\mathbf R$, and $x_v$ denotes the $v^{th}$ element of $\mathbf x$.

\subsection{Check-Table $\mathbb{T}$}

Note that for one channel realization, both $\mathbf R$ and $\chi$ are independent of time. In other words, to decode different received symbols for one channel realization, the only term in (\ref{eq:node_pw}) which depends on time is ${\breve y}_u$. Consequently, a check-table $\mathbb{T}$ is constructed to store all values of $r_{u,v}x$, where $r_{u,v}\neq0$ and $x\in\chi$, before starting the tree search procedure. Equations (\ref{eq:node_weight}) and (\ref{eq:node_pw}) imply that only one real multiplication is needed instead of $2N-u+2$ for each node to calculate the node weight by using $\mathbb{T}$. As a result, the number of real multiplications can be significantly reduced.

Taking the square QAM lattice structure into consideration, $\chi$ can be divided into two smaller sets $\chi_{1}$ with negative elements and $\chi_{2}$ with positive elements. Take 16-QAM for example, $\chi=\left\lbrace -3,-1,1,3\right\rbrace$, then $\chi_{1}=\left\lbrace -3,-1\right\rbrace$ and $\chi_{2}=\left\lbrace 1,3\right\rbrace$. Any negative element in $\chi_{1}$ has a positive element with the same absolute value in $\chi_{2}$. Consequently, in order to build $\mathbb{T}$, only terms in the  form of $r_{u,v}x$, where $r_{u,v}\neq0$ and $x\in\chi_{1}$, need to be calculated and stored. Hence, the size of $\mathbb{T}$ is
\begin{equation}
|\mathbb{T}|=\frac{N_R|\chi|}{2}, \label{eq:size_ct}
\end{equation}
where $N_R$ denotes the number of non-zero elements in matrix $\mathbf{R}$, and $|\chi|$ denotes the size of $\chi$.

In order to build $\mathbb{T}$, both the number of terms that need to be stored and the number of real multiplications required are $|\mathbb{T}|$. Since the channel is assumed to be flat fading and $\mathbb{T}$ only depends on the $\mathbf R$ matrix and $\chi$,
only one $\mathbb{T}$ needs to be built in one burst. If the burst length is very long, its computational complexity can be neglected.

\subsection{Intermediate Node Weights}

Define 
\begin{equation}
M({\mathbf x}^{(u)})=\breve{y}_u-\sum^{2N}_{v=u+1}{r_{u,v}x_v}, \label{eq:node_iw}
\end{equation}
with $M({\mathbf x}^{(2N)})=0$, then (\ref{eq:node_pw}) can be rewritten as
\begin{equation}
w_{pw}({\mathbf x}^{(u)})=|M({\mathbf x}^{(u)})-r_{u,u}x_u|^{2}. \label{eq:node_pw_iw}
\end{equation}

Equation (\ref{eq:node_iw}) shows that $M({\mathbf x}^{(u)})$ is independent of $x_u$, which means for any node not in the last level of the search tree, all children nodes share the same $M({\mathbf x}^{(u)})$. In other words, for these nodes, the $M({\mathbf x}^{(u)})$ values need to be calculated only once to get the whole set of weights for their children nodes. Consequently, the number of operations will be reduced if $M({\mathbf x}^{(u)})$ values are stored at each node, except nodes of the last level, until the whole set of their children are visited. Based on (\ref{eq:node_weight}), (\ref{eq:node_iw}), and (\ref{eq:node_pw_iw}), by storing the $M({\mathbf x}^{(u)})$ values, the number of real additions needed to get all partial weights of the children nodes at layer $u$, for a parent node of layer $u+1$, reduces to $2N-u+|\chi|$ from $(2N-u+1)|\chi|$. Note that after implementing the check-table $\mathbb{T}$, storing $M({\mathbf x}^{(u+1)})$ values does not affect the number of real multiplications.

\subsection{New Lattice Representation}

In our previous work \cite{Azzam_SD_NLR}, \cite{Azzam_SD_RLR}, a new lattice representation was proposed for (\ref{eq:io_real}) that enables decoding the real and imaginary parts of each complex symbol independently. Also, a near ML decoding algorithm, which combines DFS, K-best decoding, and quantization, was introduced. In this work, a different application of the lattice representation, which achieves no performance degradation, is employed.

For the new lattice representation, (\ref{eq:receive_c2r})-(\ref{eq:noise_c2r}) become
\begin{equation}
{\mathbf y}=
\begin{bmatrix}
\Re \left\lbrace \tilde{y}_1 \right\rbrace & \Im \left\lbrace \tilde{y}_1 \right\rbrace & \cdots & \Re \left\lbrace \tilde{y}_N \right\rbrace & \Im \left\lbrace \tilde{y}_N \right\rbrace
\end{bmatrix}^{T},
\label{eq:receive_nrr}
\end{equation}
\begin{equation}
{\mathbf H}=
\setlength{\arraycolsep}{0.0em}
\small
\begin{bmatrix}
\Re \left\lbrace \tilde{H}_{1,1} \right\rbrace & -\Im \left\lbrace \tilde{H}_{1,1} \right\rbrace & \cdots & \Re \left\lbrace \tilde{H}_{1,N} \right\rbrace & -\Im \left\lbrace \tilde{H}_{1,N} \right\rbrace \\
\Im \left\lbrace \tilde{H}_{1,1} \right\rbrace & \Re \left\lbrace \tilde{H}_{1,1} \right\rbrace & \cdots & \Im \left\lbrace \tilde{H}_{1,N} \right\rbrace & \Re \left\lbrace \tilde{H}_{1,N} \right\rbrace \\
\vdots &  \vdots & \ddots & \vdots & \vdots \\
\Re \left\lbrace \tilde{H}_{N,1} \right\rbrace & -\Im \left\lbrace \tilde{H}_{N,1} \right\rbrace & \cdots & \Re \left\lbrace \tilde{H}_{N,N} \right\rbrace & -\Im \left\lbrace \tilde{H}_{N,N} \right\rbrace \\
\Im \left\lbrace \tilde{H}_{N,1} \right\rbrace & \Re \left\lbrace \tilde{H}_{N,1} \right\rbrace & \cdots & \Im \left\lbrace \tilde{H}_{N,N} \right\rbrace & \Re \left\lbrace \tilde{H}_{N,N} \right\rbrace \\
\end{bmatrix},
\normalsize
\label{eq:channel_nrr}
\end{equation}
\begin{equation}
{\mathbf x}=
\begin{bmatrix}
\Re \left\lbrace \tilde{x}_1 \right\rbrace & \Im \left\lbrace \tilde{x}_1 \right\rbrace & \cdots & \Re \left\lbrace \tilde{x}_N \right\rbrace & \Im \left\lbrace \tilde{x}_N \right\rbrace
\end{bmatrix}^{T},
\label{eq:transmit_nrr}
\end{equation}
\begin{equation}
{\mathbf n}=
\begin{bmatrix}
\Re \left\lbrace \tilde{n}_1 \right\rbrace & \Im \left\lbrace \tilde{n}_1 \right\rbrace & \cdots & \Re \left\lbrace \tilde{n}_N \right\rbrace & \Im \left\lbrace \tilde{n}_N \right\rbrace
\end{bmatrix}^{T}.
\label{eq:noise_nrr}
\end{equation}
Define each pair of column in (\ref{eq:channel_nrr}) as one set starting from the left hand side. Then, it is obvious that the columns in each set are orthogonal, and this property has a substantial effect on the structure of the problem. Using this channel representation changes the order of the detection of the transmitted symbols. For example, the first and second levels of the search tree correspond to the real and imaginary parts of $\tilde{x}_N$, unlike the conventional SD, where these levels correspond to the imaginary parts of $\tilde{x}_N$ and $\tilde{x}_{N-1}$, respectively. The structure of the new lattice representation (\ref{eq:receive_nrr})-(\ref{eq:noise_nrr}) becomes advantageous after applying the QR decomposition to $\mathbf H$, which is formalized in the following theorem. 

\begin{theorem}
Applying QR decomposition to the real representation of the channel matrix $\mathbf{H}$, which has the aforementioned orthogonal property between the two columns in one set, produces an upper triangular matrix $\mathbf{R}$ whose elements $r_{u,u+1}$ are all zero for $u=1,3,\ldots,2N-1$.
\end{theorem}
  

\begin{proof}
Let $\mathbf{h}_u$ denote the $u^{th}$ column of $\mathbf{H}$ for $u=1, \ldots, 2N$. Then define $\mathbf{f}_1 = \mathbf{h}_{1}$, and $\mathbf{f}_u = \mathbf{h}_u - \sum_{v=1}^{u-1} \mathrm{\phi}_{\mathbf{f}_v}(\mathbf{h}_u)$ for $u=2, \ldots, 2N$, based on the Gram-Schmidt algorithm, where $\mathrm{\phi}_{\mathbf{f}_v}(\mathbf{h}_u)$ is the projection of vector $\mathbf{h}_u$ onto
$\mathbf{f}_v$ defined by
\begin{align}
\mathbf{\phi}_{\mathbf{f}_v}(\mathbf{h}_u) = {\langle \mathbf{h}_u, \mathbf{f}_v \rangle \over \langle \mathbf{f}_v, \mathbf{f}_v \rangle} \mathbf{f}_v.
\end{align}
Also define $\mathbf{e}_u = {\mathbf{f}_u \over \| \mathbf{f}_u \|}$ for $u=1, \ldots, 2N$. Then the column vectors of the channel matrix $\mathbf{H}$ can be rewritten as \\
$\mathbf{h}_1 = \mathbf{e}_1 \| \mathbf{f}_1 \|$, \\
$\mathbf{h}_2 = \mathrm{\phi}_{\mathbf{f}_1}(\mathbf{h}_2) + \mathbf{e}_2 \| \mathbf{f}_2 \|$, \\
$\mathbf{h}_3 = \mathrm{\phi}_{\mathbf{f}_1}(\mathbf{h}_3) + \mathrm{\phi}_{\mathbf{f}_2}(\mathbf{h}_3) + \mathbf{e}_3 \|\mathbf{f}_3 \|$, \\
$\vdots$ \\
$\mathbf{h}_u = \sum_{v=1}^{u-1} \mathrm{\phi}_{\mathbf{f}_v}(\mathbf{h}_u) + \mathbf{e}_u \| \mathbf{f}_u \|$. \\
Then, define $\mathbf{Q} = \left[\mathbf{e}_1 \, \cdots \, \mathbf{e}_{2N}\right]$, and these equations can be written in matrix form as 
\begin{align}
\mathbf{Q} \left[
\begin{array} {c c c c c}
\| \mathbf{f}_1 \| & \langle \mathbf{e}_1, \mathbf{h}_2 \rangle & \langle\mathbf{e}_1, \mathbf{h}_3 \rangle  & \ldots \\
0 & \| \mathbf{f}_2 \| & \langle \mathbf{e}_2, \mathbf{h}_3 \rangle & \ldots \\
0 & 0 & \|\mathbf{f}_3 \| & \ldots \\
\vdots & \vdots & \vdots & \ddots
\end{array}
\right].
\end{align}
Obviously, the matrix $\mathbf{Q}$ is unitary, and the matrix on the right is the upper triangular
$\mathbf{R}$ matrix of the QR decomposition.

Now the goal is to show that the terms $\langle \mathbf{e}_u, \mathbf{h}_{u+1} \rangle$ are zero for $u=1, 3, \ldots, 2N-1$. Three observations conclude the proof.

First, since $\mathbf{h}_u$ and $\mathbf{h}_{u+1}$ are orthogonal for $u=1, 3, \ldots, 2N-1$, then $\mathrm{\phi}_{\mathbf{f}_u}(\mathbf{h}_{u+1}) = \mathrm{\phi}_{\mathbf{f}_{u+1}}(\mathbf{h}_u) = 0$ for the same $u$.

Second, the inner products of $\mathbf{f}_v$ for $v=1, 3, \ldots, u-2$ with the columns $\mathbf{h}_{u}$ and $\mathbf{h}_{u+1}$ are equal to the inner products of $\mathbf{f}_{v+1}$ with the columns $\mathbf{h}_{u+1}$ and $-\mathbf{h}_{u}$ respectively, which are formalized as
\begin{align*}
\langle \mathbf{f}_v, \mathbf{h}_u \rangle = \langle \mathbf{f}_{v+1}, \mathbf{h}_{u+1} \rangle, \\
\langle \mathbf{f}_v, \mathbf{h}_{u+1} \rangle = -\langle \mathbf{f}_{v+1}, \mathbf{h}_u \rangle,
\end{align*}
for $u=1, 3, \ldots, 2N-1$ and $v=1,3,\ldots,u-2$. These properties become obvious by using the first observation and
revisiting the special structure of (\ref{eq:channel_nrr}).

Third, making use of the first two observations, and noting that $\|\mathbf{h}_u \| = \| \mathbf{h}_{u+1} \|$
for $u=1, 3, \ldots, 2N-1$, it can be easily shown that $\| \mathbf{f}_u \| = \| \mathbf{f}_{u+1} \|$
for the same $u$. \\
Then,
\begin{align*}
\langle \mathbf{e}_u, \mathbf{h}_{u+1} \rangle = & \langle {\mathbf{f}_u \over \| \mathbf{f}_u \|}, \mathbf{h}_{u+1} \rangle \\
= & {1 \over \| \mathbf{f}_u \|} \langle \mathbf{h}_u - \sum_{v=1}^{u-1} \mathrm{\phi}_{\mathbf{f}_v}(\mathbf{h}_u), \mathbf{h}_{u+1} \rangle \\
= & {1 \over \| \mathbf{f}_u \|} (\langle \mathbf{h}_u, \mathbf{h}_{u+1} \rangle - {\langle \mathbf{h}_u, \mathbf{f}_1 \rangle \langle \mathbf{f}_1, \mathbf{h}_{u+1} \rangle \over \langle \mathbf{f}_1, \mathbf{f}_1 \rangle} \\
& - {\langle \mathbf{h}_u, \mathbf{f}_2 \rangle \langle \mathbf{f}_2, \mathbf{h}_{u+1} \rangle \over \langle \mathbf{f}_2, \mathbf{f}_2 \rangle} - \cdots \\ 
& - {\langle \mathbf{h}_u, \mathbf{f}_{u-2} \rangle \langle {\mathbf{f}_{u-2}}, \mathbf{h}_{u+1} \rangle \over \langle \mathbf{f}_{u-2}, \mathbf{f}_{u-2} \rangle} \\
& - {\langle \mathbf{h}_{u}, \mathbf{f}_{u-1} \rangle \langle \mathbf{f}_{u-1}, \mathbf{h}_{u+1} \rangle \over \langle \mathbf{f}_{u-1}, \mathbf{f}_{u-1} \rangle}).
\end{align*}
Now, applying the above observations, then
\begin{align*}
\langle \mathbf{e}_u, \mathbf{h}_{u+1} \rangle = & {1 \over \| \mathbf{f}_u \|} (0 - {\langle \mathbf{h}_u, \mathbf{f}_1 \rangle \langle \mathbf{f}_1, \mathbf{h}_{u+1} \rangle \over \| \mathbf{f}_1 \|^2} \\
& - { -\langle \mathbf{f}_1, \mathbf{h}_{u+1} \rangle \langle \mathbf{h}_u, \mathbf{f}_1 \rangle \over \| \mathbf{f}_1 \|^2} -  \cdots \\ 
& - {\langle \mathbf{h}_u, \mathbf{f}_{u-2} \rangle \langle \mathbf{f}_{u-2}, \mathbf{h}_{u+1} \rangle \over \| \mathbf{f}_{u-2} \|^2} \\ 
& - { -\langle \mathbf{f}_{u-2}, \mathbf{h}_{u+1} \rangle \langle \mathbf{h}_u, \mathbf{f}_{u-2} \rangle \over \| \mathbf{f}_{u-2} \|^2}) \\ 
= & 0.
\end{align*}
This concludes the proof.
\end{proof}

The locations of these zeros introduce orthogonality between the real and imaginary parts of every detected symbol, which can be taken advantage of to reduce the computational complexity of SD. The following example is provided to explain this fact.

\textsl{Example:} Consider a MIMO system with $N_r=N_t=N=2$, and employing $4$-QAM. Then, SD constructs a tree with $2N=4$ levels, where the branches coming out from each node represent the real values in the set $\chi=\lbrace-1,1\rbrace$. This tree is shown in Fig. \ref{fig:tree_eg}. Now using the real-valued lattice representation (\ref{eq:receive_nrr})-(\ref{eq:noise_nrr}), and applying the QR decomposition to the channel matrix, the input-output relation is given by 

\begin{equation}
\begin{bmatrix}
\breve{y}_1 \\ \breve{y}_2 \\ \breve{y}_3 \\ \breve{y}_4  
\end{bmatrix}
=
\begin{bmatrix}
r_{1,1} & 0 & r_{1,3} & r_{1,4} \\
0 & r_{2,2} & r_{2,3} & r_{2,4} \\
0 & 0 & r_{3,3} & 0 \\
0 & 0 & 0 & r_{4,4}
\end{bmatrix}
\begin{bmatrix}
x_{1} \\ x_{2} \\ x_{3} \\ x_{4}  
\end{bmatrix}
+
\begin{bmatrix}
\breve{n}_{1} \\ \breve{n}_{2} \\ \breve{n}_{3} \\ \breve{n}_{4}  
\end{bmatrix},
\label{eq:io_qr_eg}
\end{equation}
where $[\breve{n}_{1} \, \breve{n}_{2} \, \breve{n}_{3} \, \breve{n}_{4}]^T = \breve{\mathbf n} = \mathbf{Q}^H \mathbf{n}$.

\ifCLASSOPTIONonecolumn
\begin{figure}[!p]
\centering
\includegraphics[width=1.0\textwidth]{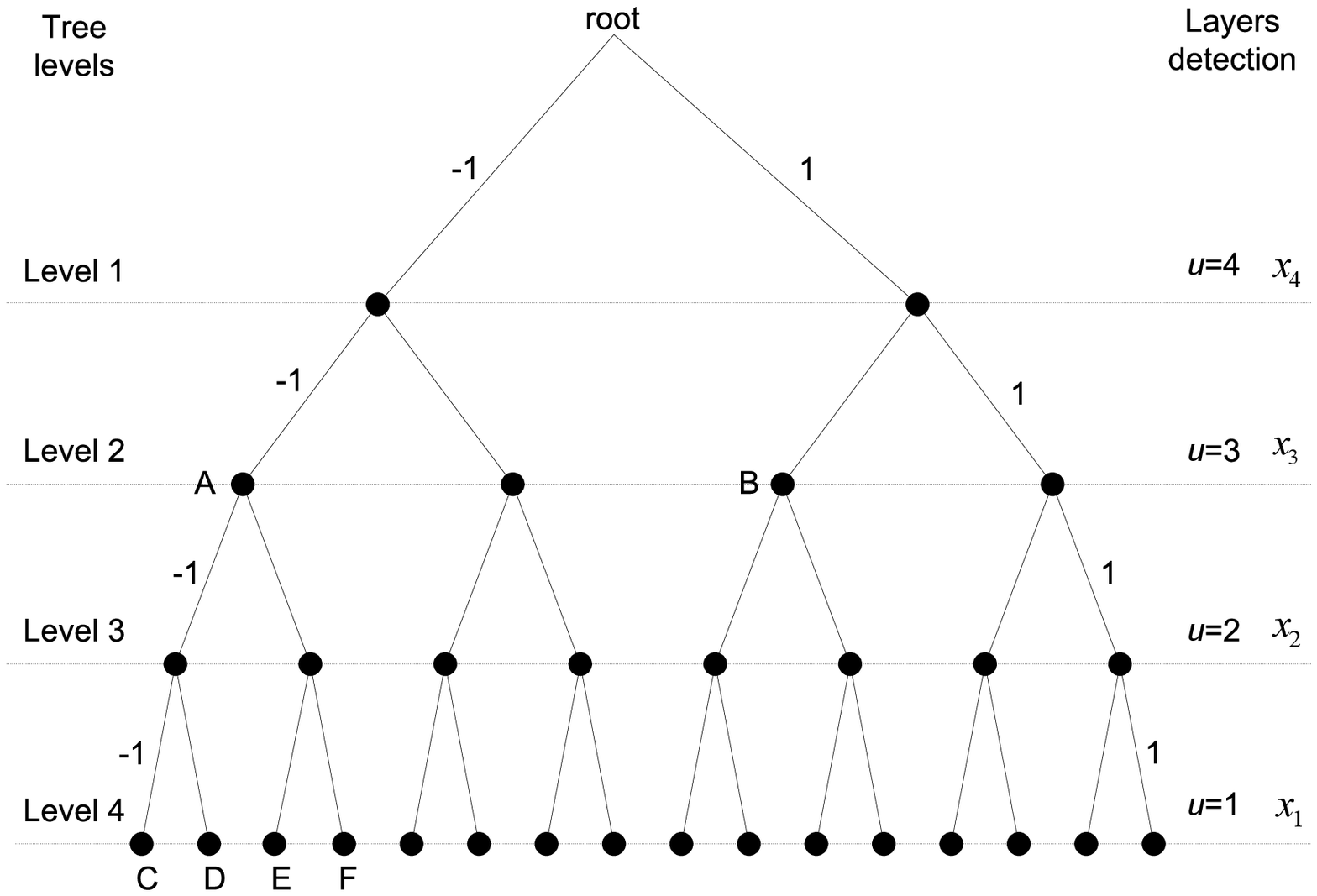} 
\caption{Tree structure for a $2 \times 2$ system employing $4$-QAM.}
\label{fig:tree_eg}
\end{figure}
\else
\begin{figure}[!t]
\centering
\includegraphics[width=0.5\textwidth]{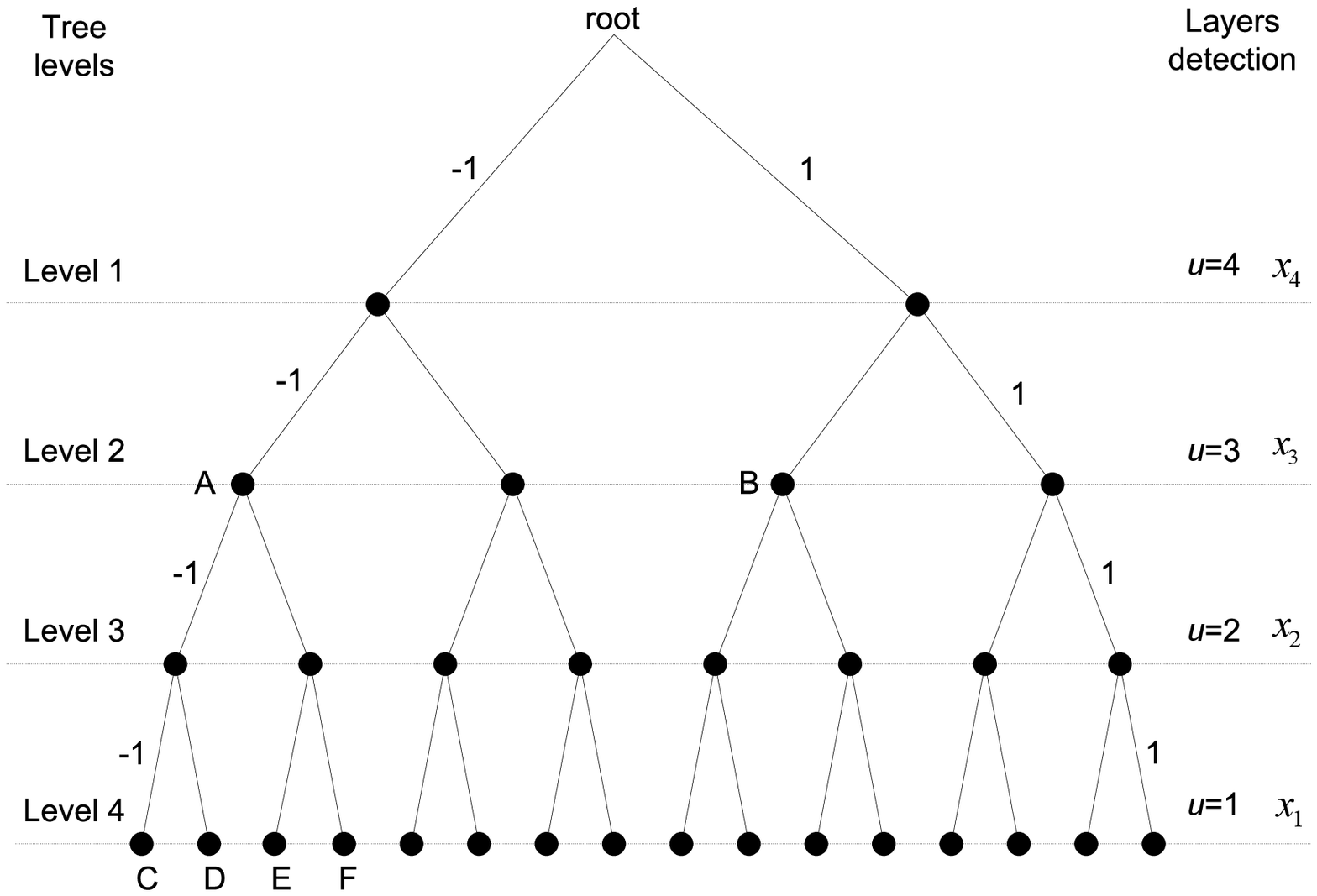} \caption{Tree structure for a $2 \times 2$ system employing $4$-QAM.}
\label{fig:tree_eg}
\end{figure}
\fi

Based on (\ref{eq:node_pw}) and (\ref{eq:io_qr_eg}), calculating partial node weights for the first level and the second level are independent, similar to the third level and the fourth level, because of the additional zero locations in the $\mathbf R$ matrix. For instance, the partial weights of node $A$ and node $B$ only depend on $x_3$ but $x_4$, and the partial weights of node $C$, node $D$, node $E$, and node $F$, depend on $x_4$, $x_3$, and $x_1$ (not on $x_2$). In other words, the partial weights of node $A$ and node $B$ are equal, and only need to be calculated once. Similarly, partial weights of node $C$ and node $D$ can be used when calculating the partial weights of node $E$ and node $F$, respectively. 

SD is then modified because of this feature. Once the tree is searched in layer $u$, where $u$ is an odd number, partial weights of this node and all of its sibling nodes are computed, stored, and recycled when calculating partial node weights with the same grandparent node of layer $u+2$ but with different parent nodes of layer $u+1$.         

By applying the modification, further complexity reduction is achieved beyond the reduction due to the check-table $\mathbb{T}$ and intermediate $M({\mathbf x}^{(u+1)})$ values. For a node of layer $u+2$, where $u$ is an odd number, let $\alpha\in[0,|\chi|]$ denote the number of non-pruned branches for its children nodes of layer $u+1$. If $\alpha=0$, which means all branches of its children nodes of layer $u+1$ are pruned, the number of operations needed stay the same. If $\alpha\neq0$, to get all partial weights of its grandchildren nodes in layer $l$, the number of real multiplications and real additions reduce further from $(\alpha+1)|\chi|$ to $2|\chi|$, and $(\alpha+1)(2N-u-1+|\chi|)+\alpha$ to $2(2N-u-1+|\chi|)$, respectively. 

%% file: Soft.tex
\section{Soft Bit Metric} \label{sec:Soft}

In many MIMO systems, channel coding, e.g., convolutional coding, is employed to provide coding gain. In this section, an application of SD employing a proposed smart implementation with very low complexity for calculating the soft bit metrics of a bit-interleaved convolutional-coded MIMO system is presented. Bit-interleaved convolutional-coded MIMO systems exist in many standards, such as $802.11$n or WiMax.

Define $\mathbf{c}$ as the codeword generated by a convolutional encoder with rate $R_c$ from the information bits. The codeword $\mathbf{c}$ is first interleaved by a bit-wise interleaver, then the interleaved bits are modulated by Gray mapped square QAM and transmitted through the $N$ transmit antennas. At the $k^{th}$ time instant, an $N \times 1$ complex-valued symbol vector $\tilde{\mathbf{x}}_k$ is transmitted at the transmitter and an $N \times 1$ complex-valued symbol vector $\tilde{\mathbf{y}}_k$ is received at the receiver. The location of the coded bit $c_{k'}$ within the complex-valued symbol sequence $\tilde{\mathbf{X}} = [\tilde{\mathbf{x}}_1 \, \cdots \, \tilde{\mathbf{x}}_K]$ is given as $k' \rightarrow (k, l, i)$, where $k$, $l$, and $i$ are the time instant in $\tilde{\mathbf{X}}$, the symbol position in $\tilde{\mathbf{x}}_k^\prime$, and the bit position on the label of the scalar symbol $\tilde{x}_{k,l}^\prime$, respectively. At the receiver, instead of decoding each transmitted symbol as in (\ref{eq:mld}), the ML soft bit metrics are calculated for each coded bit as
\begin{align}
\gamma^{l,i}(\tilde{\mathbf{y}}_{k}, c_{k'}) = \min_{\tilde{\mathbf{x}}^ \in \tilde{\xi}_{c_{k'}}^{l,i}} \| \tilde{\mathbf{y}}_{k} - \tilde{\mathbf{H}} \tilde{\mathbf{x}} \|^2, \label{eq:ML_bit_metrics}
\end{align}
where $\tilde{\xi}_{c_{k'}}^{l,i}$ is a subset of $\tilde{\chi}^N$, defined as
\begin{align*}
\tilde{\xi}_{b}^{l,i} = \{ \tilde{\mathbf{x}} = [\tilde{x}_1 \, \cdots \, \tilde{x}_N ]^T : \tilde{x}_{u|u=l} \in \tilde{\chi}_{b}^{i}, \textrm{ and } \tilde{x}_{u|u \neq l} \in \tilde{\chi}\},
\end{align*}
and $\tilde{\chi}_{b}^{i}$ denotes a subset of $\tilde{\chi}$ whose labels have $b \in \{0, 1\}$ in the $i^{th}$ bit position. Finally, the ML decoder, which uses Viterbi decoding, makes decisions according to the rule
\begin{align}
\mathbf{\hat{c}} = \arg\min_{\mathbf{c}} \sum_{k'} \gamma^{l,i}(\tilde{\mathbf{y}}_{k}, c_{k'}).
\label{eq:Decision_Rule}
\end{align}

SD can be employed to solve (\ref{eq:ML_bit_metrics}). Define $\mathbf{y}_k$ and $\mathbf{x}_k$ as the corresponding real-valued representations of $\tilde{\mathbf{y}}_k$ and $\tilde{\mathbf{x}}_k$, respectively. For square QAM with size $2^m$ where $m$ is an even integer, the first and the remaining $\frac{m}{2}$ bits of labels for the $2^m$-QAM are generally Gray coded separately as two $2^{\frac{m}{2}}$-PAM constellations, and represent the real and the imaginary axes, respectively. Assume that the same Gray coded mapping scheme is used for the the real and the imaginary axes. As a result, each element of $\mathbf{x}_k$ belongs to a real-valued signal set $\chi$, and one bit in the label of $\mathbf{x}_k$ corresponds to $c_{k^\prime}$. The new position of $c_{k^\prime}$ in the label of $\mathbf{x}_k$ needs to be acquired as $k^\prime\rightarrow(k,\hat{l},\hat{i})$, which means $c_{k^\prime}$ lies in the $\hat{i}^{th}$ bit position of the label for the $\hat{l}^{th}$ element of real-valued vector symbol $\mathbf{x}_k$. Let $\chi_b^{\hat{i}}$ denote a subset of $\chi$ whose labels have $b \in \{0, 1\}$ in the $\hat{i}^{th}$ bit position. Define $\xi_{c_{k'}}^{\hat{l},\hat{i}} \subset \chi^{2N}$ as 
\begin{align*}
\xi_{b}^{\hat{l},\hat{i}} = \{ \mathbf{x} = [x_1 \, \cdots \, x_{2N} ]^T : x_{u|u=\hat{l}} \in \chi_{b}^{\hat{i}}, \textrm{ and } x_{u|u \neq \hat{l}} \in \chi\}.
\end{align*}
Then SD solves 
\begin{equation}
\gamma^{l,i}(\tilde{\mathbf{y}}_k, c_{k^\prime}) = \min\limits_{\mathbf{x} \in \Omega} \| \breve{\mathbf{y}}_k - \mathbf{R} \mathbf{x} \|^2 \label{eq:sd_bit_metrics}
\end{equation}
with $\breve{\mathbf{y}}_k=\mathbf{Q}^H \mathbf{y}_k$, where $\Omega \subset \xi_{c_{k'}}^{\hat{l},\hat{i}}$, and $\Vert \breve{\mathbf{y}}_{k} - \mathbf{R} \mathbf{x} \Vert^2<\delta^{2}$.
The only difference between (\ref{eq:sd_bit_metrics}) and (\ref{eq:sd}) is that the search space is now $\xi_{c_{k'}}^{\hat{l},\hat{i}}$ instead of $\chi^{2N}$. In other words, the problem can be viewed as a pruning algorithm on a tree of depth $2N$, whose branches correspond to elements drawn from the set $\chi$, except for branches of the layer $u=\hat{l}$, which correspond to elements drawn from the set $\chi^{\hat{i}}_{c_{k'}}$. As a result, proposed smart implementation of SD can be applied. Other than the computational complexity reduction at each node achieved by the reduced complexity SD presented in Section \ref{sec:PSD}, two more techniques are employed by the proposed smart implementation to reduce the computational complexity further.

\subsection{Acquiring Initial Radius by ZF-DFE}
The initial radius $\delta$ should be chosen properly, so that it is not too small or too large. Too small an initial radius results in too many unsuccessful searches and thus increases complexity, while too large an initial radius results in too many lattice points to be searched.

In this work, for $c_{k'}=b$ where $b \in \{0,1\}$, ZF-DFE is used to acquire an estimated real-valued vector symbol $\breve{\mathbf{x}}_k^b$, which is also the Baiba point \cite{Agrell_CPS}. Then the square of initial radius $\delta_b^2$, which guarantees no unsuccessful searches is calculated by
\begin{equation}
\delta_b^2=\Vert \breve{\mathbf{y}}_k-\mathbf{R}\breve{\mathbf{x}}_k^b \Vert^2.
\label{eq:IR_ZFDFE}
\end{equation}

The estimated real-valued vector symbol $\breve{\mathbf{x}}_k^b$ is detected successively starting from $\breve{x}_{k, 2N}^b$ until $\breve{x}_{k, 1}^b$, where $\breve{x}_{k, u}^b$ denotes the $u^{th}$ element of $\breve{\mathbf{x}}_k^b$. The decision rule on $\breve{x}_{k, u}^b$ is
\begin{equation}
\breve{x}_{k, u}^b=\left\{
\begin{array}{ll}
\arg\min\limits_{x\in\chi} {| \breve{y}_{k,u} - M(\breve{\mathbf x}^{(u)})- R_{u,u}x |},&u\neq \hat{l},\\ \arg\min\limits_{x\in\chi^{\hat{i}}_b} {| \breve{y}_{k,u} - M(\breve{\mathbf x}^{(u)}) - R_{u,u}x |},&u=\hat{l}.\\
\end{array}
\right.
\label{eq:ZFDFE} 
\end{equation}


The estimation of the symbols (\ref{eq:ZFDFE}) can be carried out recursively by rounding (or quantizing) to the nearest constellation element in $\chi$ or $\chi^{\hat{i}}_b$.

\subsection{Reducing Number of Executions in SD}

For the $k^{th}$ time instant, the real-valued vector symbol $\mathbf{x}_k$ carries $mN$ bits. Since each bit generates two bit metrics for $c_{k'}=0$ and $c_{k'}=1$, then $2mN$ bit metrics in total need to be acquired. However, some bit metrics have the same value, hence SD can be modified to be executed less than $2mN$ times, as observed in \cite{Studer_SOSD}.

Define $\hat{\mathbf{x}}_k$, $\hat{\mathbf{x}}_k^{c_{k'}}$, and $\gamma_{k}$ as
\begin{equation}
\hat{\mathbf{x}}_k = \arg\min\limits_{\mathbf{x} \in \chi^{2N}} \| \breve{\mathbf{y}}_k - \mathbf{R} \mathbf{x} \|^2,
\label{eq:ML_symbol} 
\end{equation}
\begin{equation}
\hat{\mathbf{x}}_k^{c_{k'}} = \arg\min\limits_{\mathbf{x} \in \xi_{c_{k'}}^{\hat{l},\hat{i}}} \| \breve{\mathbf{y}}_k - \mathbf{R} \mathbf{x} \|^2,
\label{eq:bit_metric_symbol_estimated} 
\end{equation}
and
\begin{equation}
\gamma_{k} = \| \breve{\mathbf{y}}_k - \mathbf{R} \hat{\mathbf{x}}_k \|^2,
\label{eq:ML_symbol_weight} 
\end{equation}
respectively. 
Then 
\begin{equation}
\gamma^{l,i}(\tilde{\mathbf{y}}_k, c_{k^\prime}) =  \| \breve{\mathbf{y}}_k - \mathbf{R} \hat{\mathbf{x}}_k^{c_{k'}} \|^2,
\label{eq:bit_metric_symbol} 
\end{equation}
Note that $\xi_0^{\hat{l},\hat{i}} \cup \xi_1^{\hat{l},\hat{i}} = \chi^{2N}$ and $\xi_0^{\hat{l},\hat{i}} \cap \xi_1^{\hat{l},\hat{i}} = \emptyset$. Then
\begin{equation}
\gamma_{k} = \min{\{\gamma^{l,i}(\tilde{\mathbf{y}}_k, c_{k^\prime}=0), \gamma^{l,i}(\tilde{\mathbf{y}}_k, c_{k^\prime}=1) \} },
\label{eq:symbol_weight_relation} 
\end{equation}
which means that, for the $mN$ bits corresponding to $\mathbf{x}_k$, the smaller bit metric for each bit of $c_{k'}=0$ and $c_{k'}=1$ have the same value $\gamma_{k}$. 

Let $\hat{b}_{\hat{i}}^{\hat{l}}\in \{0,1\}$ denotes the value of the $\hat{i}^{th}$ bit in the label of $\hat{x}_{k,\hat{l}}$, which is the $\hat{l}^{th}$ element of $\hat{\mathbf{x}}_k$. Then  
\begin{equation}
\gamma^{l,i}(\tilde{\mathbf{y}}_k, c_{k^\prime}=\hat{b}_{\hat{i}}^{\hat{l}})=  \gamma^k.
\label{eq:bit_metric_smaller} 
\end{equation}

First, two bit metrics $\gamma^{l,i}(\tilde{\mathbf{y}}_k, c_{k^\prime}=0)$ and $\gamma^{l,i}(\tilde{\mathbf{y}}_k, c_{k^\prime}=1)$ for one of the $mN$ bits corresponding to $\mathbf{x}_k$ and their related  $\hat{\mathbf{x}}_k^{c_{k'}}$ are derived by SD. Then the $\hat{\mathbf{x}}_k^{c_{k'}}$ corresponding to the smaller bit metric is chosen to be $\hat{\mathbf{x}}_k$, and $\gamma_{k}$ is acquired by (\ref{eq:symbol_weight_relation}). For each of the other $mN-1$ bits, $\gamma^{l,i}(\tilde{\mathbf{y}}_k, c_{k^\prime}=\hat{b}_{\hat{i}}^{\hat{l}})$ is acquired by (\ref{eq:bit_metric_smaller}), and $\gamma^{l,i}(\tilde{\mathbf{y}}_k, c_{k^\prime}=\bar{\hat{b}}_{\hat{i}}^{\hat{l}})$ is calculated by SD. Consequently, the execution number of SD for one time instant is reduced from $2mN$ to $mN+1$.

%% file: Results_1.tex
\section{Simulation Results} \label{sec:Results}

\subsection{Uncoded Case}
To verify the proposed technique, simulations are carried out for $2\times2$ and $4\times4$ systems using $4$-QAM and $64$-QAM. Assuming a sequence of vector symbols are transmitted, and considering multiple channel realizations for each simulation, the average number of real multiplications and real additions for decoding one transmitted vector symbol are calculated. In the figures, conventional SD is denoted by CSD and proposed SD by PSD. In the simulations, $\delta^2=2\sigma_n^2N$ is chosen as the square of initial radius. A lattice point lies inside a sphere of this radius with high probability \cite{Damen_MLD}.

Fig. \ref{fig:mul_2x2_uncoded} and Fig. \ref{fig:add_2x2_uncoded} show comparisons of the number of operations between CSD and PSD for $2\times2$ systems using $4$-QAM and $64$-QAM. For $4$-QAM, the complexity gains for the average numbers of real multiplications and real additions are around $70\%$ and $45\%$ respectively at high SNR. Corresponding numbers are $75\%$ and $40\%$ respectively at low SNR. For the $64$-QAM case, the gains increase to around $70\%$ and $65\%$ at high SNR respectively, while they are around $85\%$ and $60\%$ at low SNR respectively.

Similarly, Fig. \ref{fig:mul_4x4_uncoded} and Fig. \ref{fig:add_4x4_uncoded} show complexity comparisons using $4$-QAM and $64$-QAM for $4\times4$ systems. For $4$-QAM, gains for the average number of real multiplications and real additions are around $80\%$ and $50\%$ respectively at high SNR, while they are around $85\%$ and $45\%$ respectively at low SNR. For $64$-QAM, gains rise to around $80\%$ and $75\%$ respectively at high SNR, while they are around $90\%$ and $70\%$ respectively at low SNR.   

Simulation results show that PSD reduces the complexity significantly compared to CSD, particularly for real multiplications, which are the most expensive operations in terms of machine cycles, and the reduction becomes larger as the system dimension or the modulation alphabet size increases. An important property of our PSD is that the substantial complexity reduction achieved causes no performance degradation. The proposed technique can be combined with other techniques which reduce the number of visited nodes such as SE, and other near-optimal techniques such as K-best.

\ifCLASSOPTIONonecolumn
\begin{figure}[!m]
\centering
\scalebox{.7}{\includegraphics{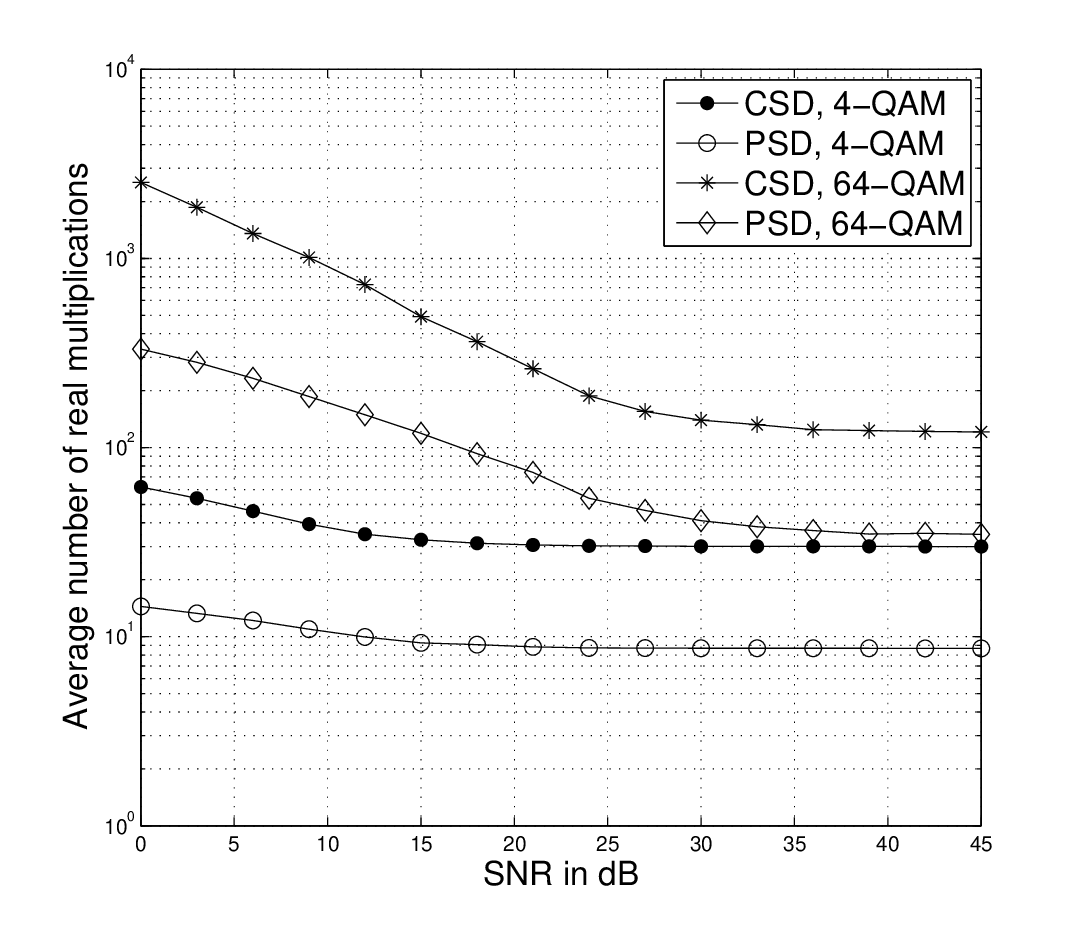}}
\caption{Average number of real multiplications vs. SNR for CSD and PSD over a $2\times2$ MIMO flat fading channel.}
\label{fig:mul_2x2_uncoded}
\end{figure}

\begin{figure}[!m]
\centering
\scalebox{.7}{\includegraphics{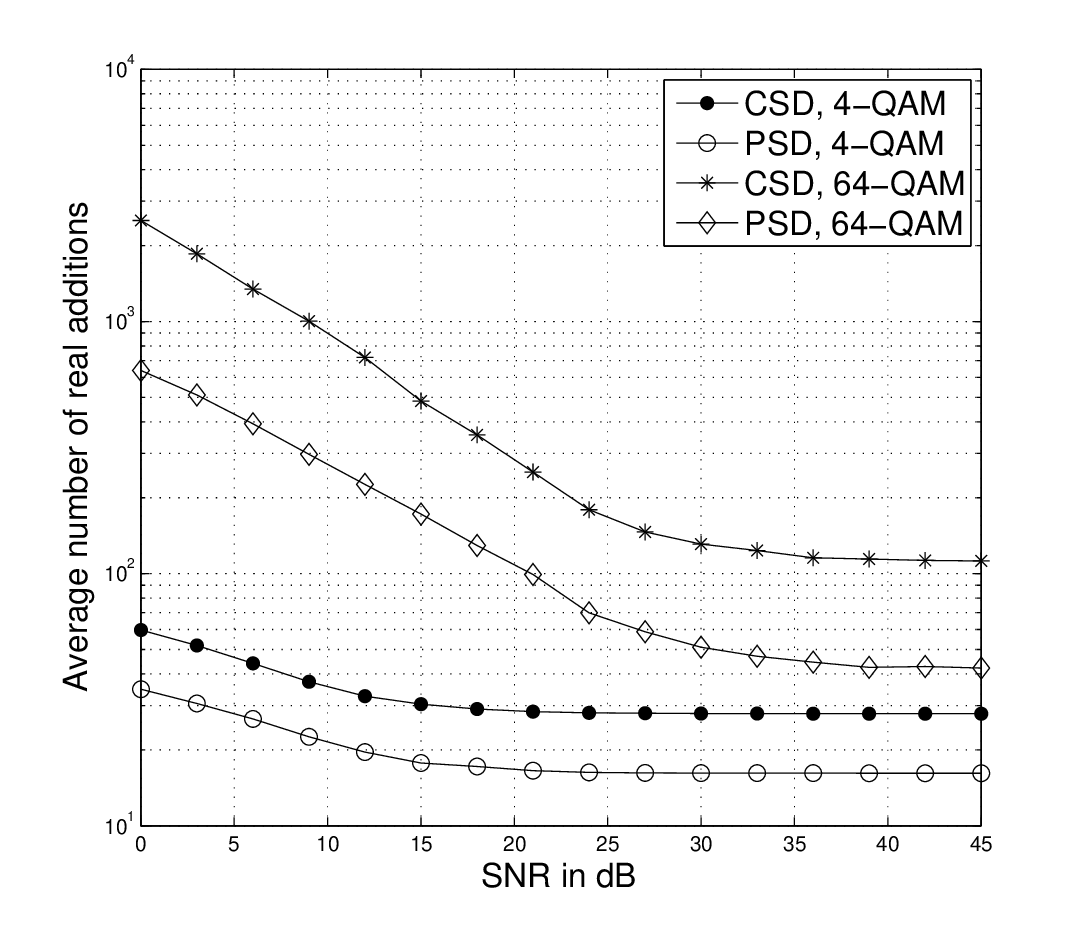}}
\caption{Average number of real additions vs. SNR for CSD and PSD over a $2\times2$ MIMO flat fading channel.}
\label{fig:add_2x2_uncoded}
\end{figure}

\begin{figure}[!m]
\centering
\scalebox{.7}{\includegraphics{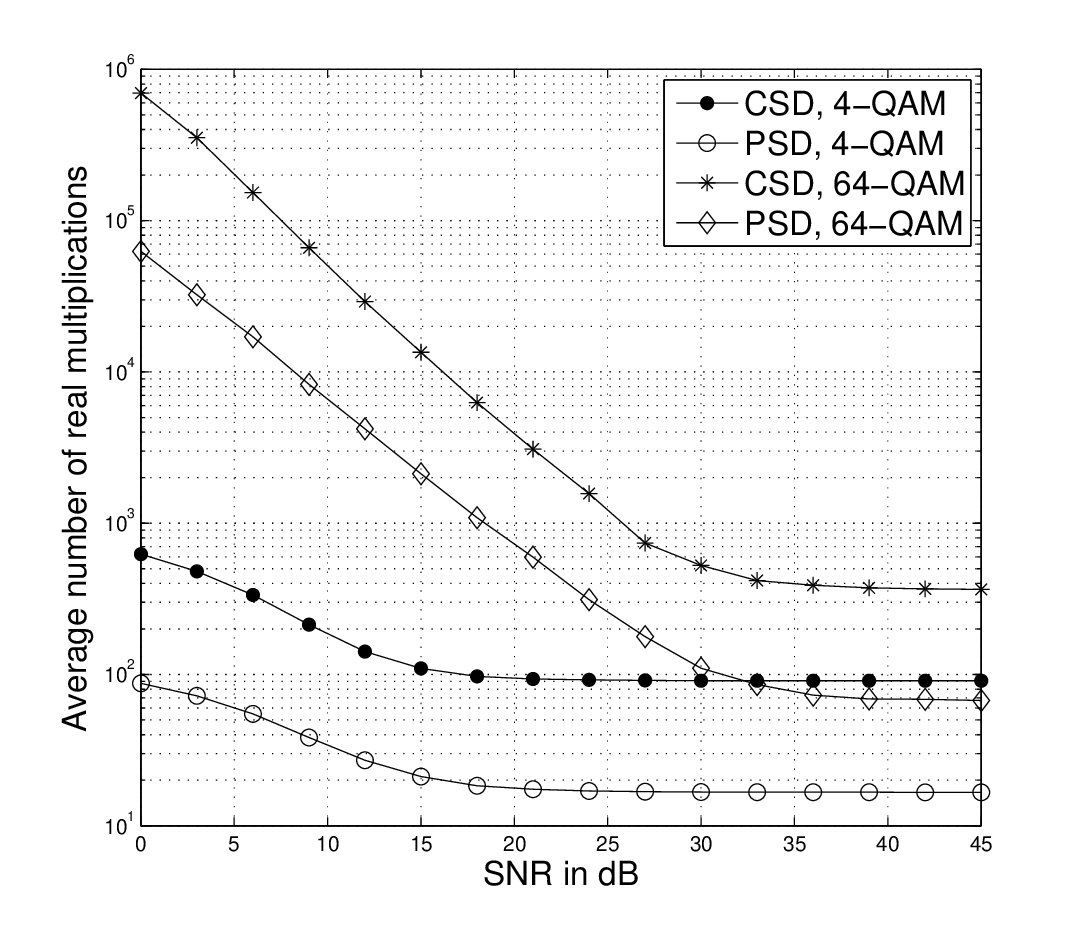}}
\caption{Average number of real multiplications vs. SNR for CSD and PSD over a $4\times4$ MIMO flat fading channel.}
\label{fig:mul_4x4_uncoded}
\end{figure}

\begin{figure}[!m]
\centering
\scalebox{.7}{\includegraphics{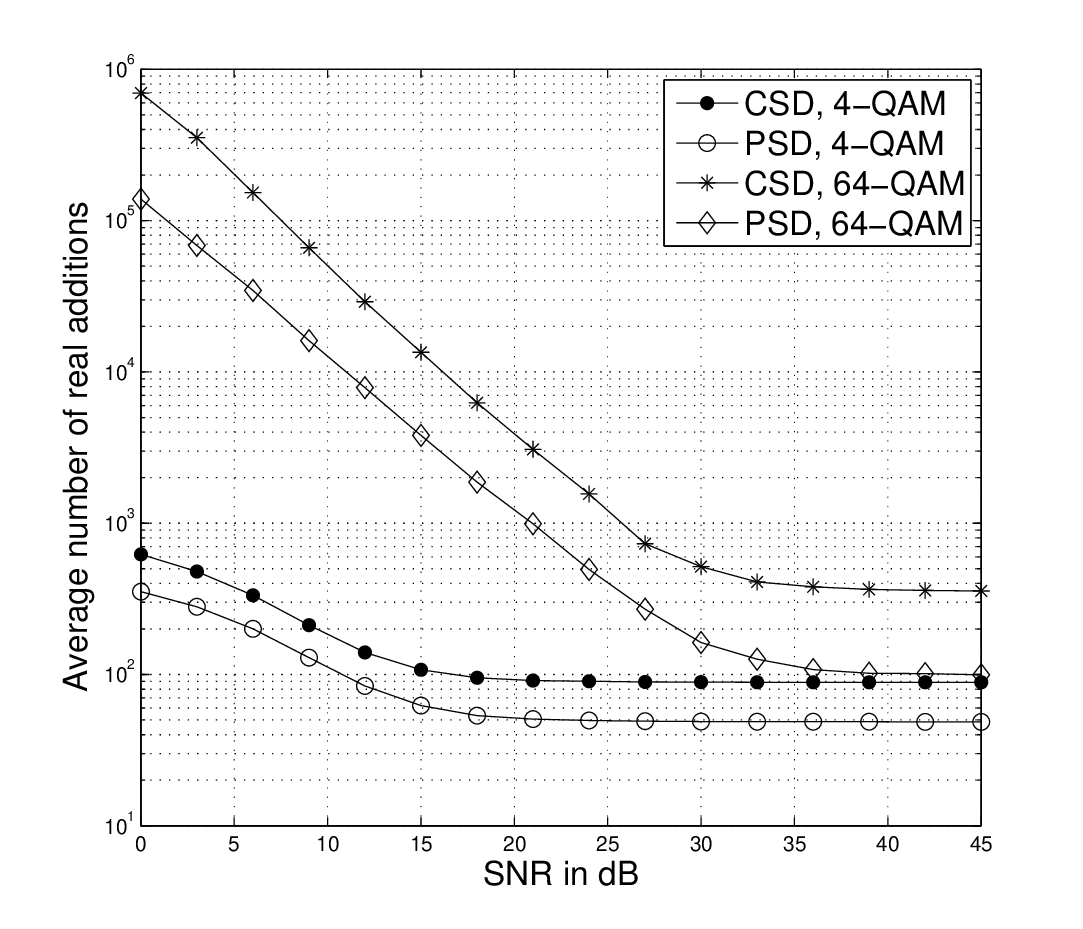}}
\caption{Average number of real additions vs. SNR for CSD and PSD over a $4\times4$ MIMO flat fading channel.}
\label{fig:add_4x4_uncoded}
\end{figure}
\else
\begin{figure}[!p]
\centering
\scalebox{.4}{\includegraphics{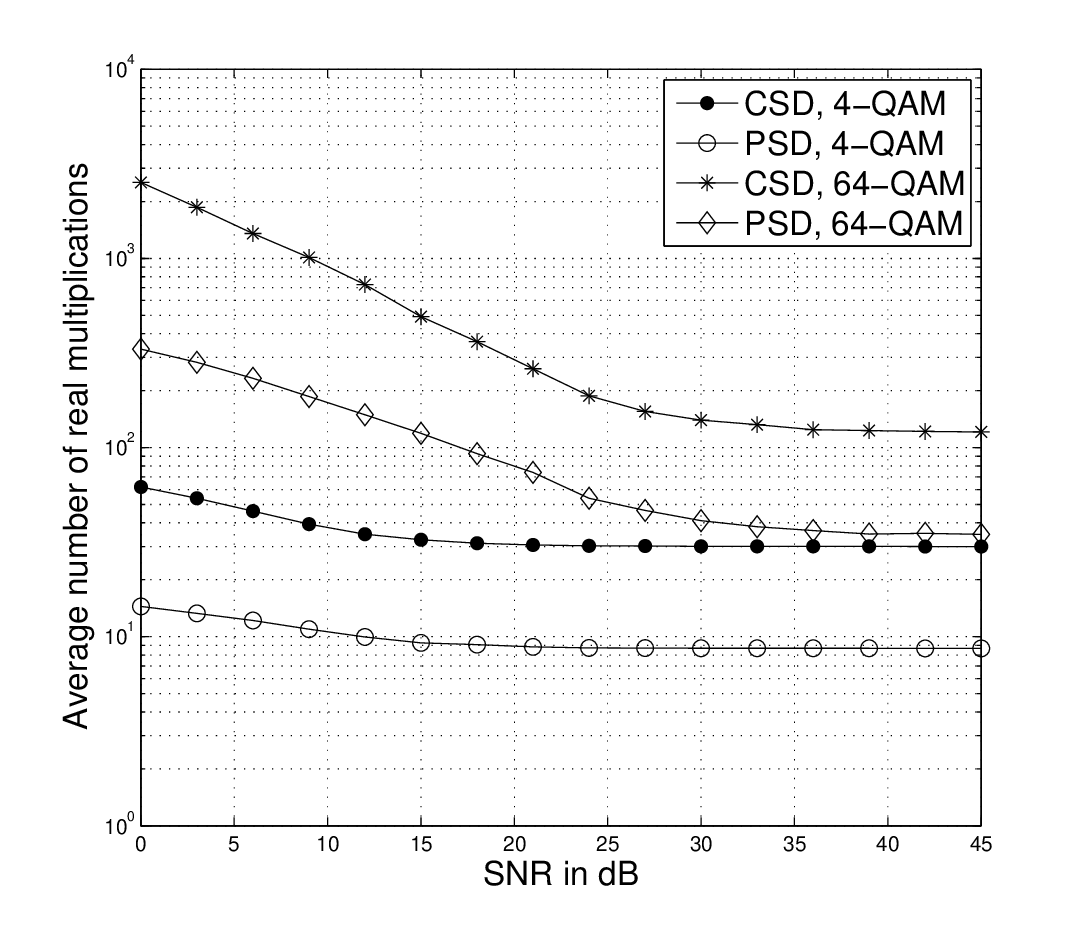}}
\caption{Average number of real multiplications vs. SNR for CSD and PSD over a $2\times2$ MIMO flat fading channel.}
\label{fig:mul_2x2_uncoded}
\end{figure}

\begin{figure}[!p]
\centering
\scalebox{.4}{\includegraphics{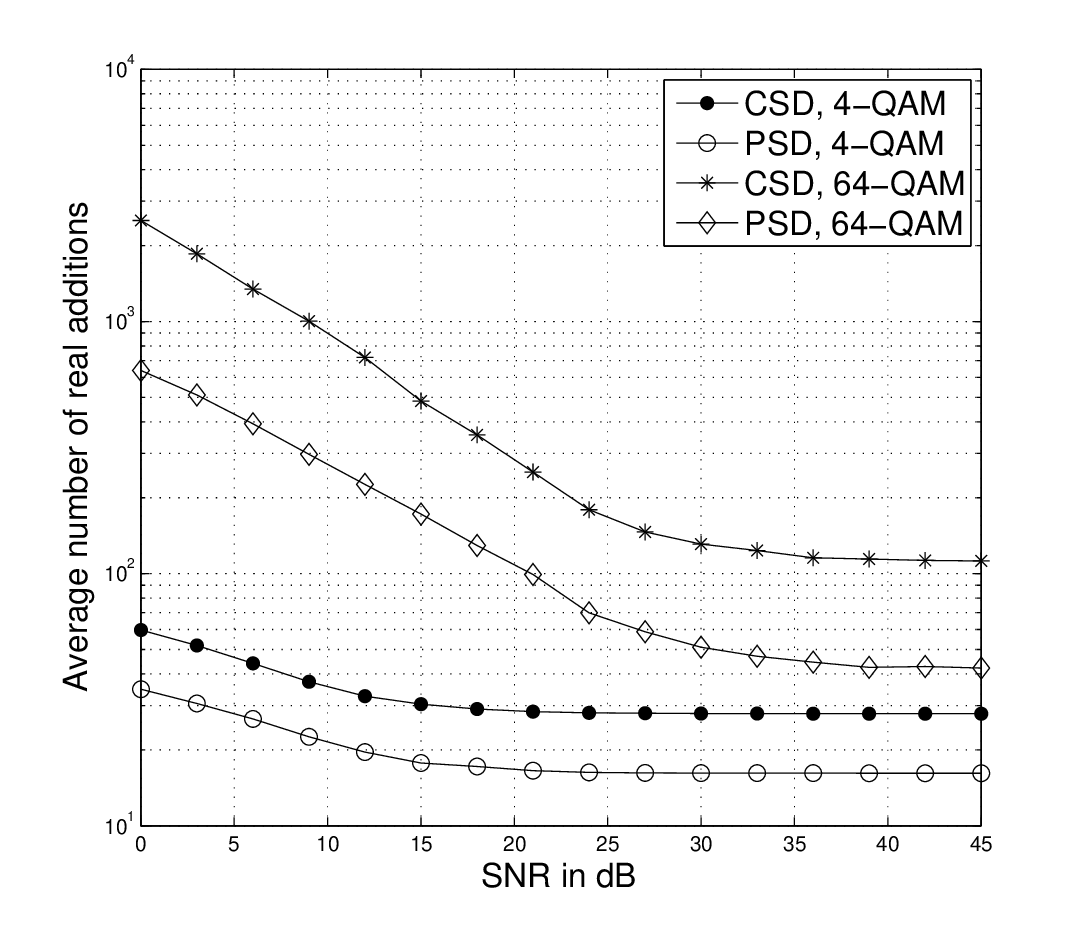}}
\caption{Average number of real additions vs. SNR for CSD and PSD over a $2\times2$ MIMO flat fading channel.}
\label{fig:add_2x2_uncoded}
\end{figure}

\begin{figure}[!p]
\centering
\scalebox{.4}{\includegraphics{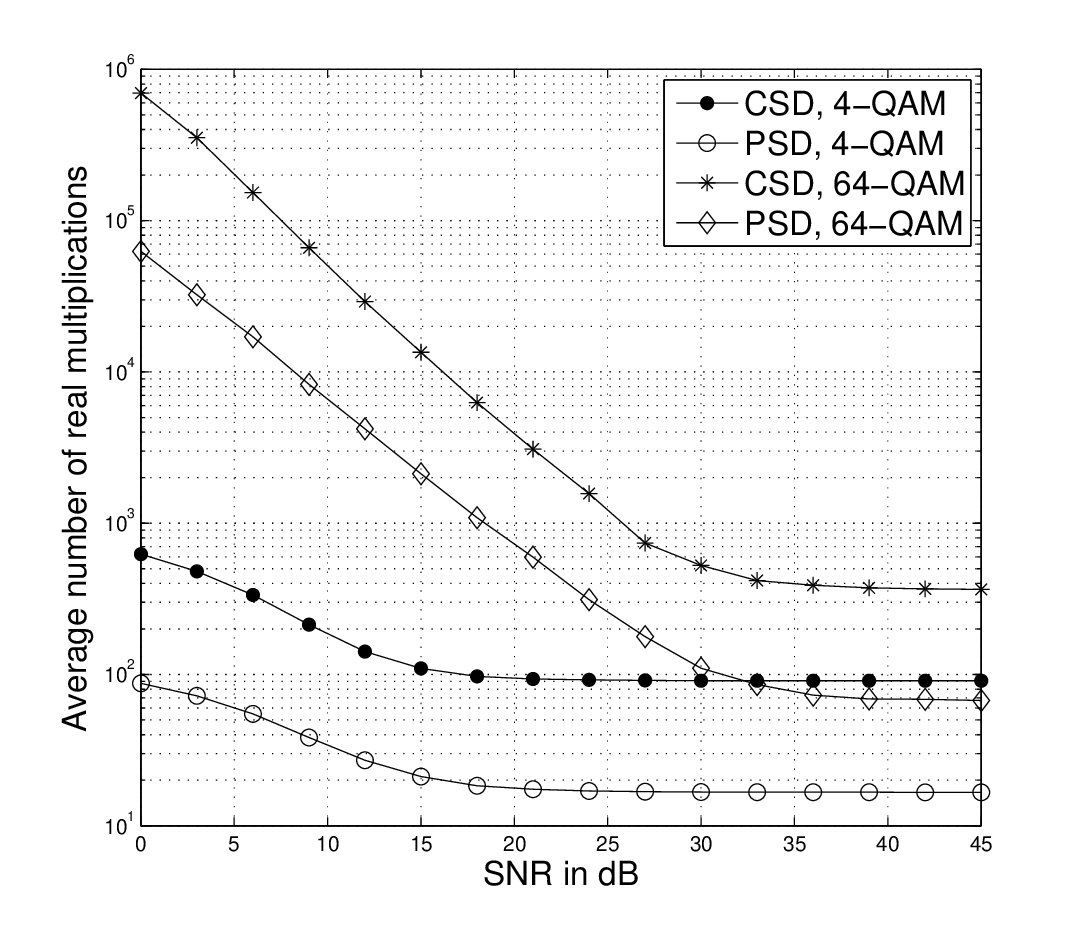}}
\caption{Average number of real multiplications vs. SNR for CSD and PSD over a $4\times4$ MIMO flat fading channel.}
\label{fig:mul_4x4_uncoded}
\end{figure}

\begin{figure}[!p]
\centering
\scalebox{.4}{\includegraphics{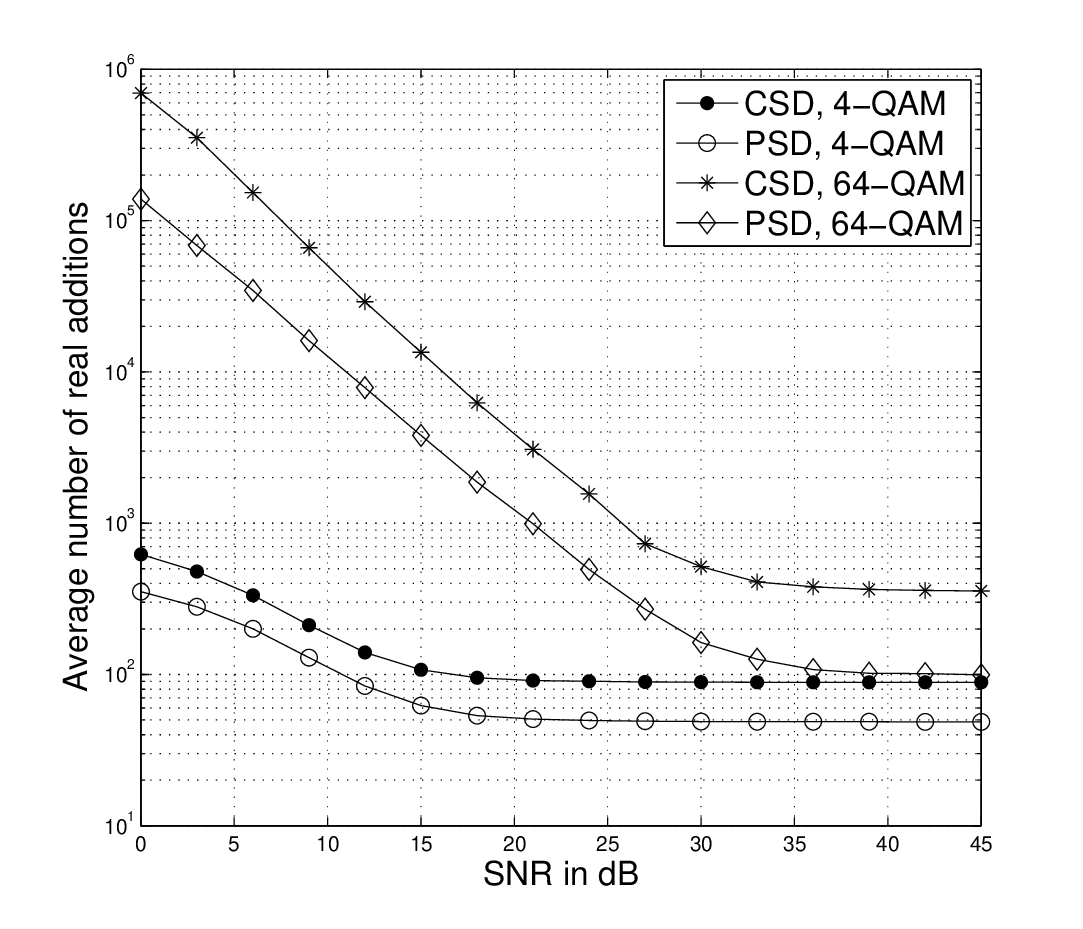}}
\caption{Average number of real additions vs. SNR for CSD and PSD over a $4\times4$ MIMO flat fading channel.}
\label{fig:add_4x4_uncoded}
\end{figure}
\fi

\subsection{Coded Case}
In \cite{park_BICMB_CP}, \cite{Park_CPMB}, and \cite{Park_MB_CP}, a novel bit-interleaved convolutional-coded MIMO technique called Bit-Interleaved Coded Multiple Beamforming with Constellation Precoding (BICMB-CP) was proposed. BICMB-CP achieves both full diversity order and full spatial multiplexing\footnotemark \footnotetext{In this paper, the term ``spatial multiplexing" is used to describe the number of spatial subchannels, as in \cite{Paulraj_ST}. Note that the term is different from ``spatial multiplexing gain" defined in \cite{Zheng_DM}.}. In \cite{Li_RC_BICMB_CP}, SD employing the proposed smart implementation was applied to reduce the computational complexity of acquiring one soft bit metric.

To verify the proposed smart implementation of SD, an $N$-dimensional implementation of Bit-Interleaved Coded Multiple Beamforming with Full Precoding (BICMB-FP), which becomes BICMB-CP when all the subchannels obtained by Singular Value Decomposition are precoded, is considered \cite{park_BICMB_CP}, \cite{Park_CPMB}, \cite{Park_MB_CP}, \cite{Li_RC_BICMB_CP}. Exhaustive Search (EXH), CSD, and Proposed Smart Implementation (PSI) are applied. The average number of real multiplications, the most expensive operations in terms of machine cycles, for acquiring one bit metric is calculated at different SNR. Since the reductions in complexity are substantial, they are expressed as orders of magnitude (in approximate terms) in the sequel.

Fig. \ref{fig:2x2} shows comparisons for $2\times2$ $R_c=\frac{2}{3}$ BICMB-FP. For $4$-QAM, the complexity of EXH is reduced by $0.4$ and $0.5$ orders of magnitude at low and high SNR respectively, by CSD. PSI yields larger reductions by $1.1$ and $1.2$ orders of magnitude at low and high SNR respectively. In the case of $64$-QAM, reductions between CSD and EXH are $1.5$ and $2.1$ orders of magnitude at low and high SNR respectively, while larger reductions of $2.6$ and $3.0$ are achieved by PSI.

Similarly, Fig. \ref{fig:4x4} shows complexity comparisons for $4\times4$ $R_c=\frac{4}{5}$ BICMB-FP. For $4$-QAM, the complexity of EXH decreases by $1.3$ and $1.5$ orders of magnitude at low and high SNR respectively. PSI gives larger reductions by $2.3$ orders of magnitude at low SNR, and $2.4$ orders of magnitude at high SNR. For the $64$-QAM case, reductions between EXH and CSD by $3.2$ and $4.4$ orders of magnitude are observed at low and high SNR respectively, while larger reductions by $4.4$ and $5.4$ are achieved by PSI.

Simulation results show that CSD reduces the complexity substantially compared to EXH, and the complexity can be further reduced significantly by PSI. The reductions become larger as the system dimension and the modulation alphabet size increase. One important property of our decoding technique which needs to be emphasized is that the substantial complexity reduction achieved causes no performance degradation. 


\ifCLASSOPTIONonecolumn
\begin{figure}[!m]
\centering
\scalebox{.7}{\includegraphics{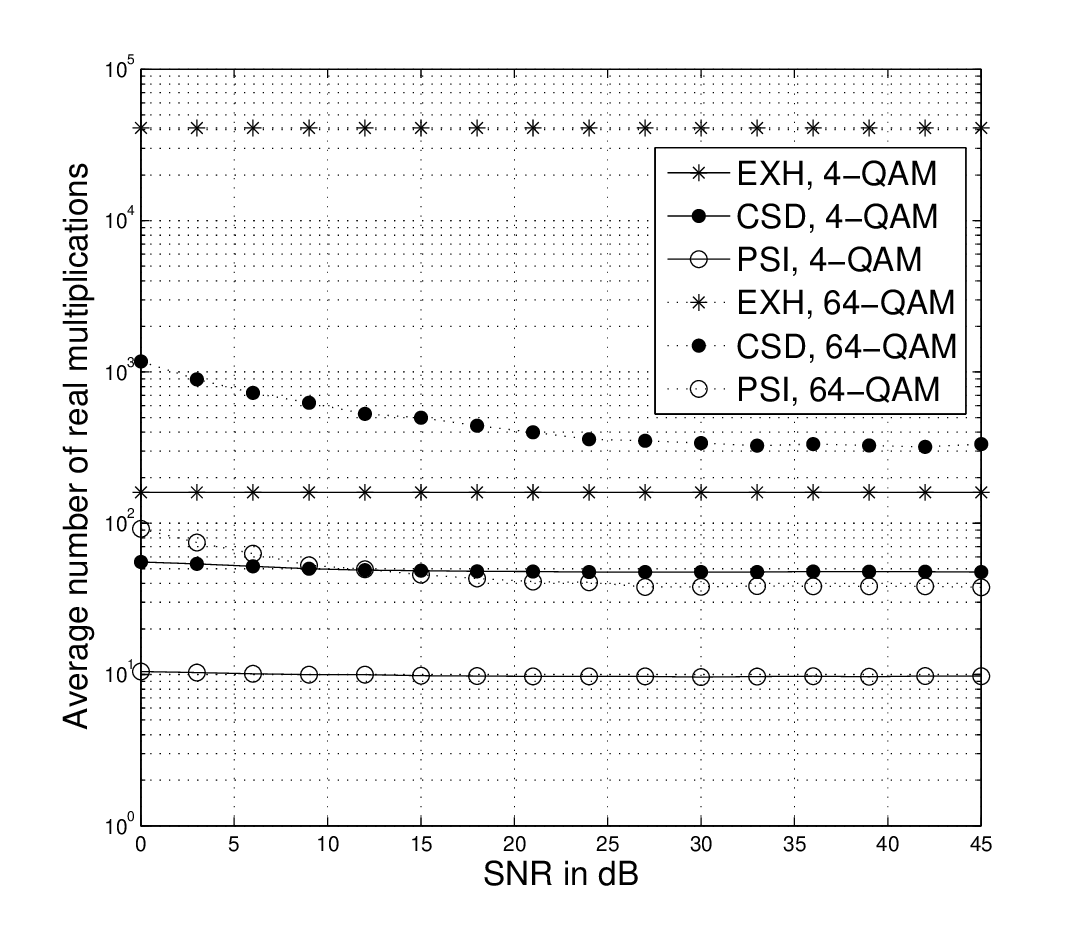}}
\caption{Average number of real multiplications vs. SNR for $2\times2$ BICMB-FP.}
\label{fig:2x2}
\end{figure}

\begin{figure}[!m]
\centering
\scalebox{.7}{\includegraphics{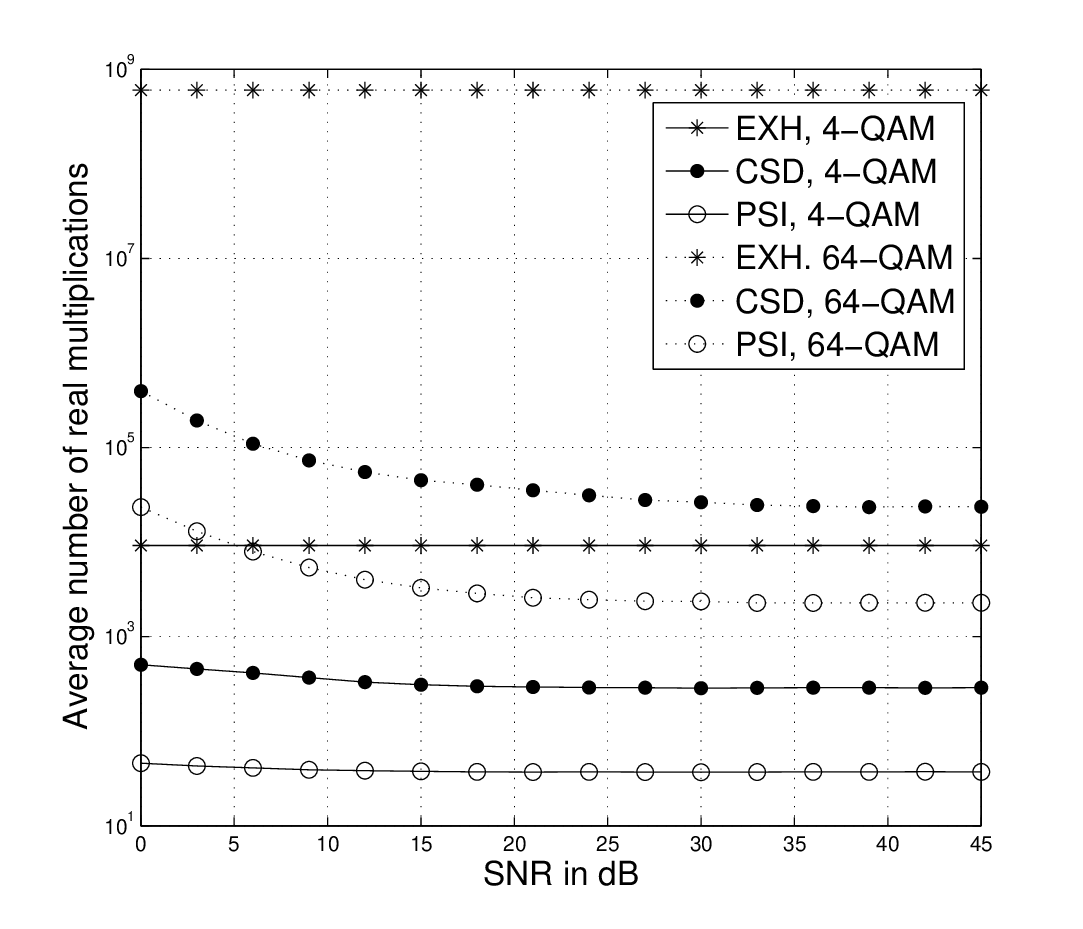}}
\caption{Average number of real multiplications vs. SNR for $4\times4$ BICMB-FP.}
\label{fig:4x4}
\end{figure}

\else
\begin{figure}[!t]
\centering
\scalebox{.4}{\includegraphics{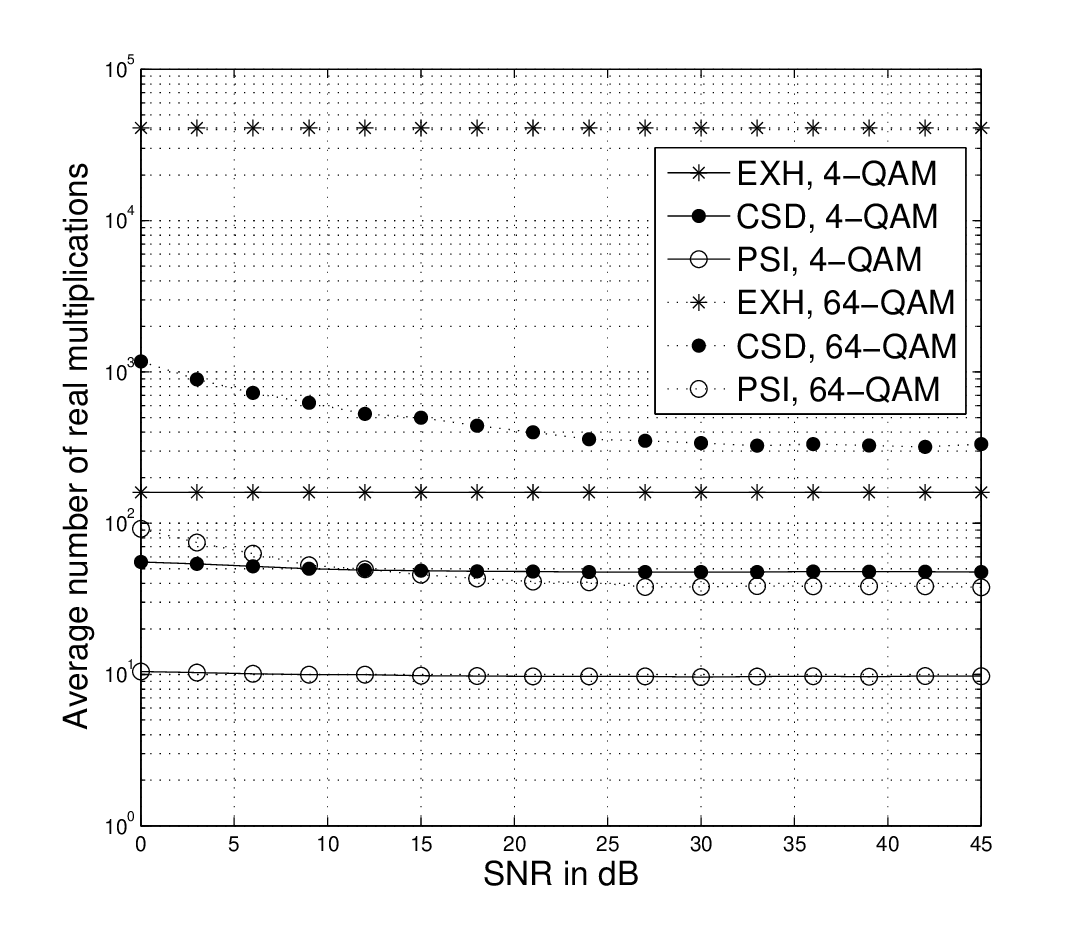}}
\caption{Average number of real multiplications vs. SNR for $2\times2$ BICMB-FP.}
\label{fig:2x2}
\end{figure}

\begin{figure}[!t]
\centering
\scalebox{.4}{\includegraphics{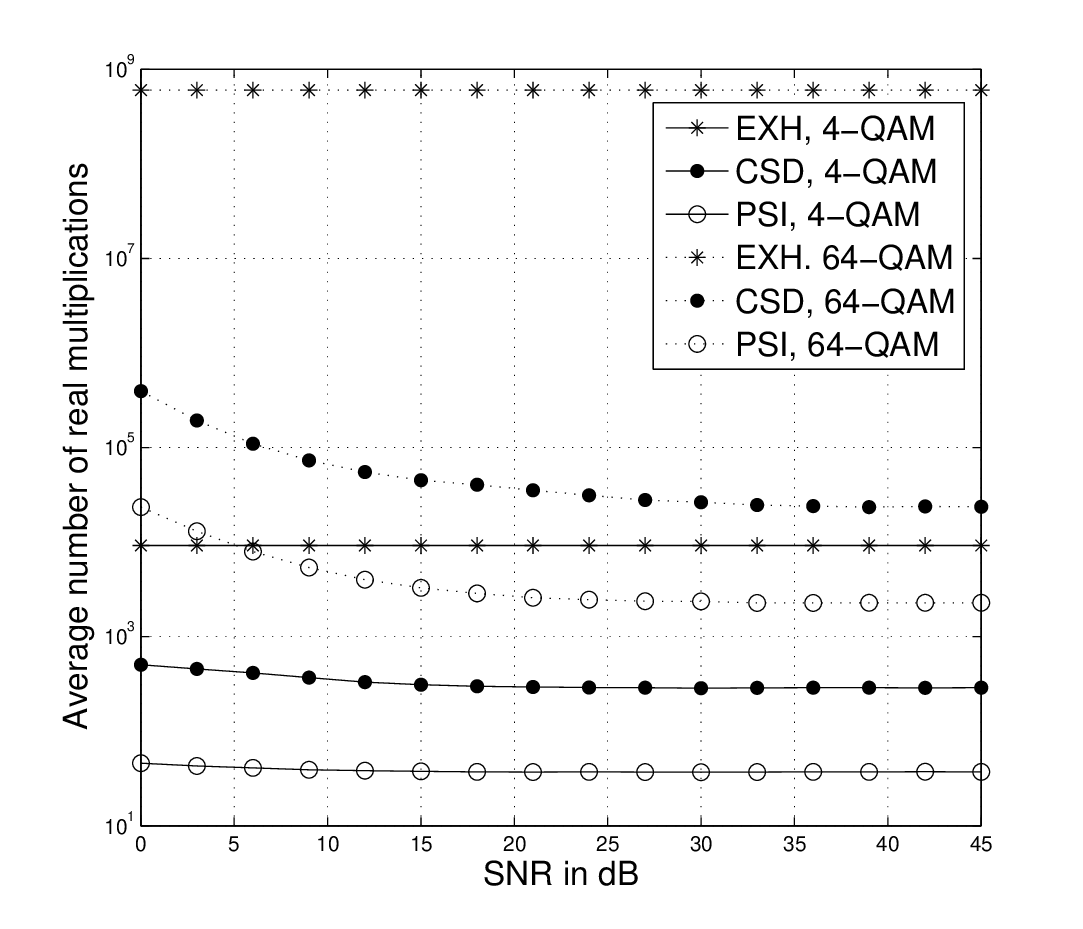}}
\caption{Average number of real multiplications vs. SNR for $4\times4$ BICMB-FP.}
\label{fig:4x4}
\end{figure}

\fi

%% file: Conclusions.tex
\section{Conclusions} \label{sec:Conclusions}

A simple and general technique to implement the SD algorithm with low computational complexity is proposed in this paper. The focus of the technique is on reducing the average number of operations required at each node for SD. The BER performance of the proposed SD is the same as conventional SD, and a substantial complexity reduction is achieved. Furthermore, an application of SD employing a proposed smart implementation with very low computational complexity for calculating the soft bit metrics of a bit-interleaved convolutional-coded MIMO system is presented. Simulation results show that these approaches achieve substantial gains in terms of the computational complexity for both uncoded and coded MIMO systems. 
